\documentclass[prb,twocolumn,,amsmath,amssymb,floatfix]{revtex4}
\usepackage{graphicx}% Include figure files
\usepackage{bm}% bold math
\usepackage[usenames]{color}
\usepackage{soul}
\sethlcolor{yellow}

\newcommand{\be}{\begin{equation}}
\newcommand{\ee}{\end{equation}}

\newcommand{\ts}[1]{\mbox{\tiny #1}}

\def \be{\begin{equation}}
\def \ee{\end{equation}}
\def \ba{\begin{array}}
\def \ea{\end{array}}
\def \bea{\begin{eqnarray}}
\def \eea{\end{eqnarray}}
\def \nn{\nonumber}

\def \half{{1\over 2}}
\def \etal{{\it {et al}}}

\def \cA{{\cal A}}

\def \a{{\alpha}}
\def \t{{\theta}}

\def \d{{\delta}}
\def \w{{\omega}}

\def \f{{\varphi}}

\def \h{{\eta}}

\def \nd{{^{\vphantom{\dagger}}}}
\def \yd{^\dagger}
\def \av#1{{\langle#1\rangle}}
\def \ket#1{{\,|\,#1\,\rangle\,}}
\def \bra#1{{\,\langle\,#1\,|\,}}

\begin{document}

\title{Intrinsic dephasing in one dimensional ultracold atom interferometers}

\author{Rafi Bistritzer and Ehud Altman}
\affiliation {Department of Condensed Matter Physics, The Weizmann
Institute of Science Rehovot, 76100, Israel}

\begin{abstract}

Quantum phase fluctuations play a crucial role in low dimensional systems.
In particular they prevent true long range phase order
from forming in one dimensional condensates, even at
zero temperature. Nevertheless, by dynamically splitting the condensate
into two parallel decoupled tubes, a macroscopic relative phase, can
be imposed on the system. This kind of setup has been proposed as a matter
wave interferometer, which relies on the interference between the displaced condensates
as a measure of the relative phase between them\cite{prentiss,jorg}.
Here we show how the quantum phase fluctuations, which are so effective in equilibrium,
act to destroy the macroscopic relative phase that was imposed as a non equilibrium initial condition
of the interferometer.
We show that the phase coherence between the
two condensates decays exponentially with a dephasing time that depends
on intrinsic parameters: the dimensionless interaction strength, sound velocity and density.
Interestingly, at low temperatures the dephasing time is almost independent of temperature.
At temperatures higher than a crossover scale $T^\star$ dephasing gains significant
temperature dependance.
In contrast to the usual phase diffusion in three dimensional traps\cite{diffusion},
which is essentially an effect of confinement,
the dephasing due to fluctuations in one dimension is a bulk effect that survives
the thermodynamic limit.

\end{abstract}

\date{\today}

\maketitle
\section{Introduction}
The existence of a macroscopic phase facilitates observation of
interference phenomena in Bose-Einstein condensates (BEC). For
example, an interference pattern involving a macroscopic number of
particles arises when a pair of condensates is let to expand freely
until the two clouds spatially overlap. Since the pioneering
experiment by Andrews \etal\cite{ketterle}, which demonstrated this
effect, it has been a long standing goal to construct matter wave
interferometers based on ultra cold atomic gases. Such devices have
promising applications in precision measurement\cite{kasevich}, and
quantum information processing\cite{jorgQI}, as well as fundamental
study of correlated quantum matter.
The key requirement for interferometric measurement
is deterministic control over the  phase difference between the two
condensates. That is, the relative phase must be well defined and
evolve with time under the sole influence of the external forces which are the
subject of measurement. This requirement was not met for example in
the setup of Ref. [\onlinecite{ketterle}], where the two condensates were essentially
independent and the relative phase between them was initially undetermined.

\begin{figure*}[t]
  \centerline{\includegraphics[width=0.9\linewidth]{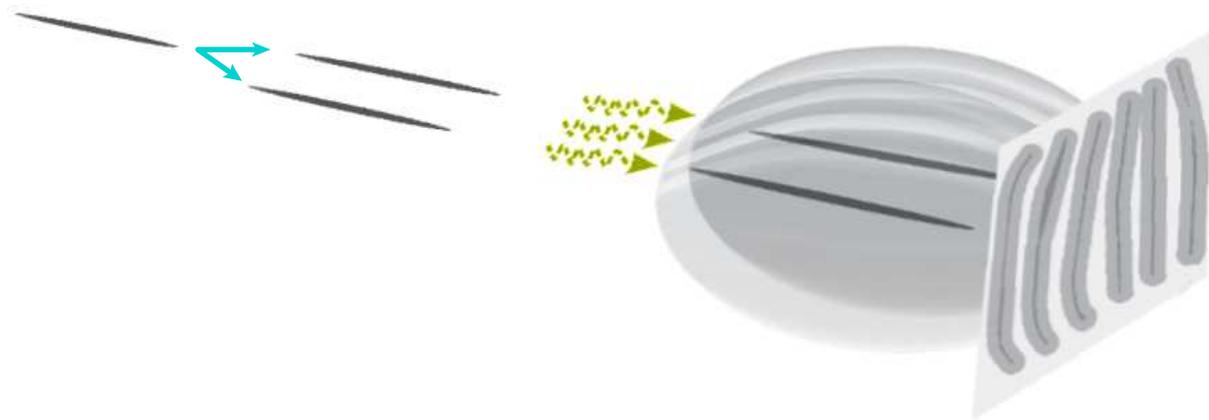}}
  \caption{Interferometer setup. At time $t=0$ a single one-dimensional condensate is split into two
  coherent parts by raising a high potential barrier along the condensate axis. After a wait time $t$
  the condensates are released from the trap and let to expand. After a sufficient expansion time
  the cloud is imaged by a probe beam sent along the condensate axis and the phase of the
  resulting interference pattern is recorded. This process is repeated many times in order to obtain
  a phase distribution at a number of different wait times $t$.
  \label{fig:setup}}
\end{figure*}

One way to initialize a system with a well defined phase is to
construct the analogue of a beam splitter whereby a single
condensate is dynamically split into two coherent parts, which serve
as the two ``arms'' of the
interferometer\cite{shin,prentiss,jorg,prentiss2}. In recent
experiments this was achieved by raising a potential barrier along
the axis of a quasi one-dimensional condensate\cite{jorg}. The split
is applied slowly compared to the transverse frequency of the trap,
but fast compared to the longitudinal time scales. Thus each atom is
transferred in this process to the symmetric superposition between
the two traps without significantly changing its longitudinal
position. An illustration of the procedure is given in Fig.
\ref{fig:setup}. In repeated experiments the condensates are
released from the trap to probe the interference at varying times
after the split. Immediately following the split the two condensates
are almost perfectly in phase. However repeated measurements at
longer times show that the phase distribution becomes gradually
broader until it becomes uniformly distributed on the interval
$[0,2\pi]$. The accuracy of interferometric measurements is limited
by such dephasing.

%This issue was addressed in the Ref. \cite{jorg} by
%repeating the experiment many times and recording the probability distribution of
%phase measurements at varying time intervals following the split.  While the
%the average phase followed deterministic evolution
%subject to the gravitational potential difference between the two
%condensates. However, the distribution of measured phases became increasingly
%broad with time. After a characteristic time of
%a few milliseconds the phase was uniformly distributed over $2\pi$.
%Such decoherence limits the accuracy with which interferometric
%measurements can be made.
%%%%%%%%%%%%%%%%%%%%%%%%%%%%%%%%%%%%%%%%%%%%%%%%%%%%%%%
%%%%%%%%%%%%%%%%%%%%%%%%%%%%%%%%%%%%%%%%%%%%%%%%%%%%%%%
%%%%%%%%%%%%%%%%%%%%%%%%%%%%%%%%%%%%%%%%%%%%%%%%%%%%%%%

In this paper we develop a theory of dephasing in interferometers
made of one dimensional Bose gases. We assume that such systems can
be well isolated from the noisy environment, so that this source of
decoherence\cite{bruder} is set at bay. Another dephasing mechanism
that has been discussed extensively in the context of condensates in
double wells is "quantum phase
diffusion"\cite{shin,diffusion,menotti}. The uncertainty in the
particle numbers in each well brought about by the split, entails a
concomitant uncertainty in the chemical potential difference, which
leads to broadening of the global relative phase on a time scale of
$\tau_D\sim \hbar\sqrt{N}/\mu$. Here $N$ is the total particle
number and $\mu$ is the chemical potential. Note that this is
essentially a finite size effect. $\tau_D$ diverges in the proper
thermodynamic limit in which the trap frequency is taken to zero
while the density is kept constant. Here we show that dephasing in
quasi one dimensional systems is dominated by a new mechanism that
involves quantum fluctuations of the local relative phase field
rather than the global phase difference between the condensates. In
contrast to the usual phase diffusion this is a bulk mechanism that
survives the thermodynamic limit. Indeed, it is well known that in
one dimensional liquids quantum phase fluctuations are extremely
effective and prevent the formation of long range order. Here we
show how these fluctuations destroy long range order in the relative
phase, that was imposed on the system by the initial conditions. We
account for the phase fluctuations using a quantum hydrodynamic
description of the Bose liquids \cite{haldane}. The phase coherence
between the condensates is then shown to decay exponentially in
time. The dephasing time we obtain is a simple function of intrinsic
parameters: the dimensionless interaction strength, the density, and
the sound velocity. These predictions are verified and extended to
finite temperatures, using direct simulation of the dynamics in the
microscopic hamiltonian. The results are also compared to the
dephasing time measured in Ref. [\onlinecite{jorg}], and are found
to be in quantitative agreement.

It is interesting to point out the
essential difference between the problem we consider and
dephasing of single particle interference effects as
considered for example in mesoscopic electron systems.
As in our problem, phase coherence in mesoscopic systems
can be defined by an interferometric measurement\cite{heiblum}.
In a Fabry-Perot setup, oscillations of the conductivity as a function of applied
Aharonov-Bohm flux
vanish exponentially with system size, thereby defining a characteristic
dephasing length.
Because what is measured is ultimately the DC conductivity,
the problem may be recast in terms of linear response theory\cite{stern}.
By contrast, dynamic splitting of the condensate in the ultra-cold atom
interferometer takes
the system far from equilibrium and the question of phase coherence
is then essentially one of quantum dynamics. The system is prepared in
an initial state determined by the ground state of a single condensate,
which then
evolves under the influence of a completely different hamiltonian,
that of the split system. Dephasing, from this point of view, is the process
that takes the system to a new steady (or quasi-steady) state.

%The hamiltonian of the system after the split is given by:
%\be
%H=\sum_{\a=1}^2 {\hbar^2\over 2M}\int dx \partial_x\psi_a\yd(x)\partial_x\psi_\a\nd(x)
%+\half\int dx dy :\rho_\a(x)v(x-y)\rho_\a(y):
%\ee

\section{Hydrodynamic theory}
We start our analysis by considering
the hydrodynamic theory that describes low energy properties of one dimensional
Bose liquids\cite{haldane}. The hydrodynamic hamiltonian for a pair
of decoupled condensates is that of two decoupled Luttinger liquids
(we set $\hbar = 1$ throughout)
\be
H={c\over 2}\sum_{\a=1}^2\int dx \left[{K\over \pi}(\nabla\f_\a)^2
+{\pi\over K}\Pi_\a^2\right],
\label{LL}
\ee
where $c$ is the sound velocity and K is the Luttinger parameter
that determines the decay of  correlations at long distance.
$\Pi_\a$ is the density fluctuation operator conjugate to the phase $\f_\a$.
The smooth component of the Bose field operator is given by $\psi_\a\approx \sqrt{\rho}\exp(i\f_\a)$,
where $\rho$ is the average density\cite{haldane}.
%An implicit parameter in this theory
%is a short distance cutoff, set by
%the microscopic healing length $\xi_h$\cite{cazallila}.

The operator corresponding to the interference signal between the two condensates
is given by\cite{pad}:
\be
A \approx \rho\int_0^{L} dx ~e^{i(\f_1(x)-\f_2(x))},
\ee
where $L$ is the imaging length.
For a pair of decoupled condensates at equilibrium,
$\av{A}=0$ while $\av{A^2}>0$\cite{pad}. Therefore the interference pattern displayed
in repeated experiments has a finite amplitude but its phase is completely random.
On the contrary, in this work we consider a pair of condensates
that are prepared out of equilibrium, with a well defined relative phase.
In this case $\av{A}$ is expected to be non vanishing at the time of the split
and decay in time as the fluctuation in relative phase grows.

Calculation of the time evolution of $\av{A}$ is greatly simplified
by the fact that the hydrodynamic theory is quadratic.
In particular, this allows decoupling in the hamiltonian of "center of mass" and relative phase
fields $\f_+=(\f_1+\f_2)/2$ and $\f_-= \f_1-\f_2$.
In other words, the hamiltonian can be rewritten as a sum of two commuting
harmonic terms $H=H_+(\f_+,\Pi_+)+H_-(\f_-,\Pi_-)$, where $\Pi_\pm$ are the conjugate
momenta of $\f_\pm$.
Moreover, the splitting process described above ensures that the initial
state may be factorized as
$\Psi[\f_+,\f_-]=\Psi_+[\f_+]\times\Psi_-[\f_-]$.
Because during the split all atoms are simply transferred to
the symmetric superposition without changing their axial state, the wave function
$\Psi_+[\f_+]$ is identical to the wave function
of the single condensate before the split.
If this condensate is at
finite temperature $\Psi_+$ is replaced
by the appropriate density matrix. On the other hand $\Psi_-$
is determined by the splitting process and will generally be strongly
localized around $\f_-(x)=0$. In fact we can find the approximate
form of this wavefunction, which will be an important input for the
time dependant calculation.

Consider a region of size $\xi_h$, which is the length scale on which
the hydrodynamic variables are defined. Let $n_1$ and $n_2$ be the operators
corresponding to the particle number in each of the condensates
within this region.
The splitting process delocalizes the particles between the two condensates,
resulting in a roughly gaussian
distribution of the relative particle number $n_-=(n_1-n_2)/2$,
with an uncertainty of order
$\sqrt{\av{n_1+n_2}/4}=\sqrt{\rho \xi_h/2}$. The phase
$\f_-$ defined on this grain in the hydrodynamic theory
is canonically conjugate to $n_-$.
Therefore the wave function in the phase representation is approximately gaussian
with\cite{squeezed} $\av{\f(x)\f(x')}\approx \d(x-x')/(2\rho)$. That is
\be
\Psi_-[\f_-] \propto e^{-\half\rho\int dx |\f(x)|^2} = e^{-\half\rho \sum_q |\f_q|^2}.
\label{Psi_m_t0}
\ee
From here on we drop the subscript (-) where no ambiguity may arise.

We are now ready to compute $\av{A}$ which depends only on $\f_-$.
The  wave-function $\Psi_-[\f_-]$ evolves in time under the influence of
the harmonic hamiltonian
\be
H_-= {c\over 2}\sum_q \left( {2\pi\over K} |\Pi_q|^2+ {K q^2\over 2\pi} |\f_q|^2 \right).
\ee
The wavefunction remains gaussian at all times:
\be
\Psi_-[\f,\f^\star]=\prod_{q} \left({1\over 2\pi W_q}\right)^{1/4}~e^{-{|\f_q|^2\over 4W_q}}e^{-i\h_q}
\label{Psi}
\ee
where $\h_q(\f_{q},\f^\star_{q},t)$ is a pure phase and
\be
W_q(t)= \av{|\f_q|^2}={1\over 2\rho}\cos^2(cqt)+{2\pi^2\rho\sin^2(cqt)\over K^2q^2}.
\label{Wq}
\ee
The expectation value $\av{A}=\cA$ in the gaussian wave function
(\ref{Psi}) is given by
\be
\cA(t)=\rho\av{e^{i\f(x)}}=\rho e^{-{1\over 2L}\sum_{q}\av{|\f_q|^2}}=\rho e^{-g(t)}.
\label{At}
\ee

In the thermodynamic limit ($L\to\infty$) the summation in (\ref{At}) can be converted
to an integral to obtain:
\bea
g(t)&\approx&{1\over 2\pi}\int_0^{\pi/\xi_h} dq \left[{1\over 2\rho}
\cos^2(cqt)+{2\pi^2\rho\sin^2(cqt)\over  K^2q^2}\right]\nn\\
%&=& g_0+{ct\over a_0(K/\pi)^2}{1\over 2\pi}\int{\sin^2x\over x^2}dx\nn\\
&\approx&g_0+{\rho ct\over 2 (K/\pi)^2}.
\label{gt}
\eea
The last equality is valid at times $t>\xi_h/c$, when the integrals become independent
of the high momentum cutoff. In addition we require that the particle number in a healing length
$\xi_h$ is large, i.e. $\rho \xi_h\gg 1$, this is always satisfied at weak coupling $K\gg 1$.
Thus we find that the phase coherence decays exponentially:
\be
\cA(t)=\cA_0 e^{-t/\tau}            \label{At_short}
\ee
with the characteristic time
\be
\tau = {2\over c\rho}  (K/\pi)^2.       \label{tau}
\ee
Note that the dephasing time $\tau$ diverges, as it must, in the non interacting limit
($K\to\infty$ and $c\to 0$). Clearly the initial state induced by the split is
an eigenstate of non interacting particles and therefore all observables including
the coherence must be time independent in this limit.

For a finite system, the discreteness of the sum in (\ref{At}) must be taken into account at times
$t \gg L/c$. In such a case, the dominant contribution to $g(t)$ arises from the $q=0$ term. This term yields
a ballistic broadening of the phase $\av{\f^2_{q=0}}\propto t^2$,
which results in a gaussian decay of the
coherence: $\cA(t)=\cA_0\exp(-t^2/\tau_L^2)$
where $\tau_L=\sqrt{\tau L/2c}$.

It is interesting to note that one can also express the time scale
associated with the finite size as $\tau_L\propto \sqrt{N}/\mu$.
Written in this way, it is clear that the gaussian term is just the
"quantum phase diffusion" discussed in the context of double well
systems\cite{shin,diffusion}. It stems from the uncertainty relation
between the uniform component of the relative phase and the
difference in total particle numbers between the wells. The uniform
component of the interaction, which drives the ballistic broadening
of $\f_{q=0}$ scales like $1/L$, making the gaussian term irrelevant
in large systems. By contrast, the exponential dephasing due to the
internal modes is length independent. The fact that these modes
produce exponential dephasing is special to one dimensional
interferometers. A similar analysis of two
parallel planar condensates yields logarithmic divergence in (\ref{gt}),
which implies power law decay of the coherence as $\cA\propto (\xi_h/ct)^\a$. We note that in
this case $K$ is not dimensionless. The power $\a$ is non universal and depends on the
short distance cutoff. In three dimensions on the other hand the only contribution to dephasing is
the usual phase diffusion due to finite size and there is
no bulk contribution.

It is striking that the hydrodynamic theory developed in this section
predicts a dephasing time which is
independent of temperature. Temperature enters the initial condition only
through the density matrix for the symmetric degrees of freedom, whereas the phase coherence
between the condensates depends on the evolution of the relative phase. Because
the symmetric and antisymmetric fields are completely decoupled within the harmonic theory,
the temperature associated with the symmetric degrees of freedom does not affect the dephasing.
However, it is clear that the full microscopic hamiltonian contains anharmonic terms
that do couple those degrees of freedom. At equilibrium
the non linear terms are irrelevant (in the renormalization group sense) and
do not affect the asymptotic long wavelength low energy correlations.
However it is natural to question the validity of the hydrodynamic description for computing
time dependent properties. In particular, here the system is prepared out of equilibrium
in a state with extensive energy relative to the ground state. Then it is a priori unclear
that the low energy correlations given by the hydrodynamic theory are sufficient to describe the dynamics.
In the next section we shall test the predictions of the hydrodynamic theory
and obtain temperature dependent corrections to the dephasing by computing
the dynamics within a microscopic model of the twin condensates.

\section{Numerical results}

For the purpose of numerical calculations we shall consider a
lattice hamiltonian, the Bose-Hubbard model on twin
chains
\be H=-J\sum_{i}\sum_{\a=1}^2 (b\yd_{\a i} b\nd_{\a
i+1}+H.c.)+{U\over 2}\sum_{\a i} n_{\a i}(n_{\a i}-1). \label{BHM}
\ee
Here $b\yd_{\a i}$ creates a Boson on site $i$ of chain $\a$. The model (\ref{BHM})
can also describe continuum systems, such as the one in Ref.
[\onlinecite{jorg}], if the average site occupation $\bar n=\av{n_i}$ is
much less than unity.

As mentioned above, we are interested in the time evolution of the expectation value
$\cA(t)=\sum_i\av{b\yd_{1i}(t)b\nd_{2i}(t)}$. The average is taken over
the wave function or density matrix at the time of the split. We will calculate $\cA(t)$
using a semiclassical approach called the truncated Wigner approximation (TWA,
cf. Ref. [\onlinecite{polkovnikov}]).
Let us first briefly review the strictly classical approximation to the dynamics.
The usual procedure is to write the Heisenberg equations
of motion for $b\nd_i$ and $b\yd_i$ and then replace
these operators by complex classical
fields $\psi_i$ and $\psi^\star_i$.
This leads to the lattice Gross-Pitaevskii (GP) equations
\be
i{d\psi_{\a,i}\over dt}=-J(\psi_{\a,i-1}+\psi_{\a,i+1})+U|\psi_{\a,i}|^2\psi_{\a,i}.
\label{GP}
\ee
Given the initial condition $\psi_{\a,i}(0)$
one can integrate the GP equations to find the value of the fields
at any time $t$ and obtain $\cA_{cl}(t)=\sum_i\psi^\star_{1i}(t)\psi_{2i}(t)$.
If the split is fully coherent $\psi_{1i}(0)=\psi_{2i}(0)$. Since the evolution
in the two chains is described by identical equations, this equality
persists to all subsequent times. Thus in the absence of external noise sources,
the classical dynamics cannot account for dephasing.
By contrast, quantum fluctuations provide an intrinsic dephasing mechanism.

Quantum corrections modify the dynamics in two ways. First, they introduce
"quantum noise" to the initial conditions. The fields
$\psi_{\a i}$ and $\psi^\star_{\a i}$ originate from non commuting quantum
operators $b\nd_{\a i}$ and $b\yd_{\a i}$
which cannot be determined simultaneously. Therefore the unique classical initial
condition should be replaced by an ensemble of initial conditions
characterized by a quantum distribution. Second, the classical trajectories
determined by (\ref{GP}) are supplemented by additional quantum paths.

It can be shown that the leading quantum correction to the dynamics
enters through the initial conditions\cite{polkovnikov}.
The essence of TWA is to integrate the GP equations starting from
an ensemble of initial conditions, which are given quantum weights
derived from the initial density matrix.
Thus within this approximation:
\be
\cA(t)=\int {\mathcal D}\{\psi^\star_{j\a},\psi_{j\a}\}_0
\rho_{\ts{W}}(\{\psi^\star_{j\a},\psi_{j\a}\}_0)\cA_{cl}(t).
\label{TWA}
\ee
Here
\bea
\rho_{\ts{W}}(\psi^\star,\psi)&\equiv& \int d\h^\star d\h
\bra{\psi-{\h\over 2}}\rho_0\ket{\psi+{\h\over 2}}\nn\\
&&\times e^{-|\psi|^2-{1\over
4}|\h|^2}e^{\half(\h^\star\psi-\h\psi^\star)}, \label{rhow} \eea is
the Wigner representation of the initial density matrix of the
system and $\ket{\psi}$ denotes a coherent state with eigenvalue
$\psi$ of the boson annihilation operator. We note that
$\rho_{\ts{W}}$ should not be thought of as a probability
distribution, because in general it can assume negative values.

The main difficulty in applying the recipe (\ref{TWA}) to compute
$\cA(t)$ is the need to find the Wigner distribution of the split condensate
at the time of the  split.
To overcome this problem we use the following procedure. Rather than tackling a split
interacting system we start the calculation from a single {\em non-interacting} condensate,
where we have an exact expression for $\rho_{\ts{W}}$\cite{polkovnikov}. The evolution of the Wigner distribution
from a non-interacting system to an interacting one is done using the TWA
by slowly increasing the interaction constant, $U$, from zero to the desired value
(see footnote [\onlinecite{footnote}]).
The heating induced by the time dependant hamiltonian is controlled by the rate at which $U$ is increased.
The next step is the split of the condensate. It involves the
doubling of the degrees of freedom at each site. Because of the way the split is carried out,
the field $\psi_i$ of the single condensate is simply  copied
to the symmetric field $\psi_{+,i}\equiv(\psi_{1i}+\psi_{2i})/\sqrt{2}$. The new
degree of freedom $\psi_{-,i}$ is chosen at this point from a Wigner distribution
\be
\rho^{(-)}_{\ts{W}}(\{\psi^\star_-,\psi_-\}_0)= \prod_i \left({2\over \pi}\right) e^{ -2|\psi_{-,i}|^2 }
\ee
which corresponds to the vacuum of the boson $b_{-,i}=(b_{1i}-b_{2i})/\sqrt{2}$.
The calculation is continued using the dynamics of the twin condensates.
%%%%%%%%%%%%%%%%%%%%%%%%%%%%%%%%%%%%%%%%%%%%%%%%%%%%%%%
\begin{figure}[t]
  \centerline{\includegraphics[width=0.98\linewidth]{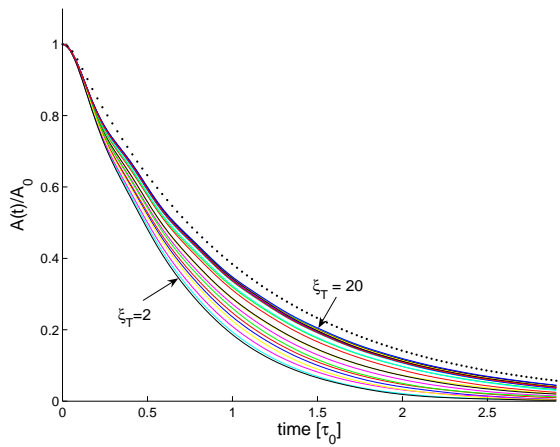}}         % decoherence
  \caption{Decay of the coherence calculated with the truncated Wigner
  approximation for $U=1$, $J=1$ and ${\bar n}=7.5$ compared to the result of
  the hydrodynamic theory (dots). The time axis is scaled to the hydrodynamic
  dephasing time $\tau_0$ (\ref{tau}).
  Lines correspond to different initial temperatures parameterized by the
  correlation length $\xi_T$ that existed before the split.
  \label{fig:coherence}}
\end{figure}
%%%%%%%%%%%%%%%%%%%%%%%%%%%%%%%%%%%%%%%%%%%%%%%%%%%%%%%
%%%%%%%%%%%%%%%%%%%%%%%%%%%%%%%%%%%%%%%%%%%%%%%%%%%%%%%
\begin{figure}[t]
  \centerline{\includegraphics[width=0.98\linewidth]{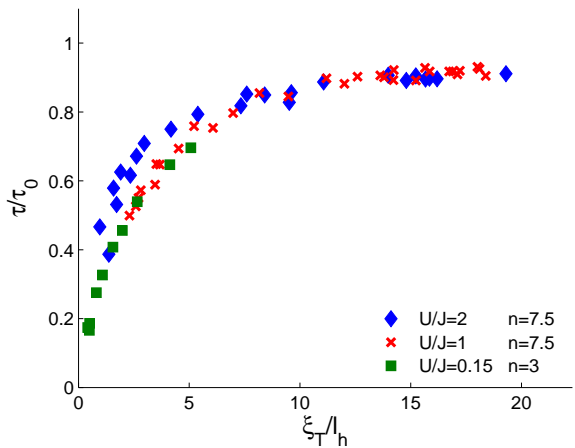}}
  \caption{Dephasing time $\tau$
  as a function of the thermal correlation length,
  which is a measure of the inverse temperature before the split.
  Data is taken in different regimes of microscopic parameters. All data is seen to collapse
  on approximately the same curve when the dephasing time is scaled to the theoretical
  time scale (\ref{tau}) and the thermal correlation length scaled by $l_h$ (see text).
  \label{fig:tauT}}
\end{figure}
%%%%%%%%%%%%%%%%%%%%%%%%%%%%%%%%%%%%%%%%%%%%%%%%%%%%%%%

The condensate (prior to the split) is prepared at various temperatures
using two different methods:
(i) By initializing the non interacting condensate in a finite temperature
density matrix.
(ii) By varying the rate by which the interactions are switched on to induce
a controlled amount of heating. The temperature of the condensate before the
split is parameterized by measuring the correlation length $\xi_T\propto 1/T$
associated with the decay of the correlation function $\av{b\yd_i b\nd_{i+r}}$.
We verified that the final result for $\cA(t)$ depends only on $\xi_T$ and not on
the method used to achieve finite temperature.

Finally, in order to compare the numerical results to the prediction of the hydrodynamic theory
we use the fact that at weak coupling there are simple relations between
the parameters of the microscopic model (\ref{BHM}) and those of the hydrodynamic theory (\ref{LL})\cite{cazalilla}.
In particular the value of the Luttinger parameter is given by
$K\approx\pi\sqrt{2J{\bar n}/U}$, and the sound velocity is $c=\sqrt{2J{\bar n} U}$.
The short distance cutoff of the hydrodynamic theory may also be extracted from (\ref{BHM}),
by finding the length over which amplitude fluctuations decay.
It is given by \cite{footnote_xi}
$l_h=\sqrt{2} \xi_h\approx a\sqrt{4J/U{\bar n}}$. Increasing $\bar n$
decreases this length scale until it reaches its lower bound which is one lattice constant $a$.
When $l_h\gg a$ the model (\ref{BHM}) in effect describes a continuum system, while for a large
site occupancy $l_h\to a$ it describes a lattice of coupled condensates.
We perform simulations in both regimes. The continuum regime is much more demanding computationally in the sense that attaining a given
thermal correlation length in this regime requires a much slower activation rate of the interaction.
This limited the temperature range we could study in the continuum regime.

The numerical results are summarized in Fig. \ref{fig:coherence} and Fig. \ref{fig:tauT}.
The dephasing time, $\tau$, is extracted by fitting $\cA(t)$ to Eq. (\ref{At_short}). At low temperatures, $\tau$ is
in good agreement with the hydrodynamic result (\ref{tau}) and shows very weak temperature dependence.
On the other hand above
a crossover temperature $T^\star$, the dephasing time gains significant
temperature dependence. The crossover scale is set by the condition $\xi_{T^\star}\sim l_h$.
This is not surprising, given that the hydrodynamic theory is expected to break down when correlations
decay over a length scale shorter than its short distance cutoff.
The agreement between the numerical results and the hydrodynamic theory at low temperatures is highly
non-trivial in view of the fact that the two approaches are based on entirely different approximations.
Most importantly, the temperature independent result (\ref{tau})
relied on the decoupling of symmetric and antisymmetric
degrees of freedom within the harmonic theory (\ref{LL}). Such decoupling does not exist in the
microscopic model (\ref{BHM}) nor in the TWA dynamics.
In the next section we shall compare the results of the
hydrodynamic theory to experimental data.

\section{Comparison with experiments}

The dephasing time (\ref{tau}) is written in terms of
parameters of the hydrodynamic theory.
In order to compare with experiments we
have to translate these parameters to
numbers that are relevant to a specific experimental situation.  For this purpose we will
consider the setup of the experiments described in Ref. [\onlinecite{jorg}].

This system consists of bosons with contact interactions
parameterized by a dimensionless interaction strength $\gamma$.
At weak coupling, $\gamma\ll 1$, the following relations hold\cite{cazalilla}:
$K=\pi/\sqrt{\gamma}$ and $c=\hbar (\rho/M)\sqrt{\gamma}$, where $M$ is the mass of an atom. Substituting these
into (\ref{tau}) we obtain:
\be
\tau=\left({2M\over \hbar \rho^2}\right){1\over \gamma^{3/2}}.
\ee
Now consider a one dimensional tube geometry with transverse trap frequency
$\w_\perp$. If the oscillator length $l_\perp$ is
much larger than the $s-wave$ scattering length $a_s$, as is the case
in Ref. [\onlinecite{jorg}],
we can use the approximate expression
$\gamma\approx 2 M\w_\perp a_s/\hbar\rho$~~\cite{olshanii,cazalilla}.
This allows us to write the dephasing time using
parameters easily determined in the experiment:
\be
\tau=\sqrt{\hbar\over 2M\rho}\left({1\over\w_\perp a_s}\right)^{3/2}
\ee
Note that  $\rho$ is the density per condensate
in the split system, that is half the density of the initial condensate.

Taking the scattering length of rubidium-87 atoms
$a_s=105 a_0$, the transverse trap frequency $\w_\perp=2\pi \times 2.1~ kHz$
and density of about $50$ atoms per micron\cite{jorg-private}, we obtain a dephasing
time $\tau\approx 4.3$ms. To estimate the dephasing time
in the experiment\cite{jorg} we use the data for the phase broadening
assuming a von Mises distribution of the phase ($f(\t)=\exp(\kappa\cos\t)/2\pi I_0(\kappa)$).
This gives approximately $2$ms, slightly shorter than our theoretical estimate.
We note that data  points marked with squares in
Fig. \ref{fig:tauT} correspond to parameters relevant to the experiment.
Unfortunately the temperature in that experiment is not known
to a good accuracy. In the future it would be interesting to
look for the universal temperature dependence implied by
the numerical results.

\section{Discussion and Conclusions}

It is well known that quantum fluctuations prevent long range phase order from forming
in one-dimensional Bose liquids. This phenomenon is most conveniently
described within the framework of
the Luttinger liquid or hydrodynamic theory.
Here we used this framework in a non equilibrium situation
to show how quantum fluctuations destroy long range
order that was imposed on the system as an initial condition. The
outcome is a simple formula describing exponential decay of
phase coherence in
interferometers made of one-dimensional condensates.
The dephasing time found in this way provides a fundamental
limit on the accuracy of such interferometers. We did not discuss
the situation in which the interferometer is prepared in a number
squeezed initial state. Such an initial condition would slow the
usual phase diffusion process\cite{menotti,prentiss2} and is
expected to similarly affect the bulk mechanism discussed in this paper.

Interestingly the dephasing time measured in
Ref. [\onlinecite{jorg}] is only slightly shorter than the prediction of the
hydrodynamic theory.
We therefore conclude that quantum phase fluctuations were probably a dominant
dephasing mechanism in that experiment.

The validity of the Luttinger liquid framework out of equilibrium is not
ensured a priori. We therefore test its predictions against numerical simulations of the microscopic
model using the Truncated Wigner approximation. At low temperatures,
the simulation results were essentially temperature independent and in good agreement with the
hydrodynamic theory. On the other hand at temperatures above a crossover scale set by the healing
length, the dephasing time displayed considerable temperature dependence.
Indeed the Luttinger liquid theory is expected to break down
above this temperature scale.

Phase fluctuations are expected to be weaker at higher dimensions.
For example three dimensional condensates can support long range order even at finite temperatures.
In this case the long range order in the relative phase imposed as an initial condition of
the interferometer
is resistant to the phase fluctuations. The situation with planar condensates is more delicate.
two dimensional systems can in principle sustain long range order at equilibrium  at zero temperature. However
the initial condition of the interferometer drives the system out of equilibrium.
The hydrodynamic theory yields power law dephasing in this case. However the power
is non universal and further work is needed to test the validity of the hydrodynamic theory.

Finally we would like to point out that the dephasing process considered in this paper
is a mechanism that brings the system, from a non equilibrium state imposed by the initial conditions
to a new steady state. According to the hydrodynamic theory this steady (or quasi-steady) state
is not yet thermal equilibrium.
In particular we find that the off diagonal correlations along the condensates
are distinctly non thermal at steady state\cite{unpublished}.
It is an interesting question whether there are processes, much slower than dephasing,
that eventually take the system toward thermal equilibrium.
This issue can be addressed by experiments.
After the phase has randomized,
correlations along the condensates can be measured
using analysis methods developed in
Refs. [\onlinecite{pad,gadp}]. Thus, we propose
that interferometric experiments
can serve as detailed
probes to address fundamental questions in non equilibrium quantum dynamics,
supplementing
measurements of global properties previously used to
touch on these issues\cite{weiss}.

{\em Acknowladgements}. We are grateful to N. Davidson, E. Demler, S. Hofferberth,
A. Polkovnikov, J. Schmiedmayer, T. Schumm, and J. H. Thywissen for useful discussions.
This work was partially supported by the U.S. Israel BSF and by an Alon fellowship.


\begin{thebibliography}{99}


\bibitem{prentiss}Y.~Shin, C.~Sanner, G.-B.~Jo, T.~A.~Pasquini,
M.~Saba, W.~Ketterle, D.~E.~Pritchard, M.~Vengalattore, and
M.~Prentiss, Phys. Rev. A {\bf 72}, 021604 (2005).

\bibitem{jorg} T. Schumm \etal, Nature Phys. {\bf 1}, 57 (2005).
\bibitem{prentiss2} G.-B. Jo \etal, cond-mat/0608585
\bibitem{diffusion} F. Sols, Physica (Amsterdam) 194B–196B, 1389 (1994);
M. Lewenstein and L. You, \prl {\bf 77}, 3489 (1996);
\bibitem{menotti} C. Menotti, J. R. Anglin, J. I. Cirac, and P. Zoller, Phys.
Rev. A 63, 023601 (2001).

\bibitem{ketterle}  M.~R.~Andrews, C.~G.~Townsend, H.~J.~Miesner,
D.~S.~Durfee, D.~M.~Kurn, and W.~Ketterle, Science {\bf 275}, 637
(1997).
\bibitem{kasevich} M. A. Kasevich, Science {\bf 298}, 1363 (2002).

\bibitem{jorgQI} M. A. Cirone, A. Negretti, T. Calarco, P. Kruger, and J. A. Schmiedmayer,
Eur. Phys. J. D 35, 165–171 (2005).

\bibitem{shin} Y. Shin \etal, Phys. Rev. Lett. 92, 050405 (2004)

\bibitem{bruder} C. Schroll, W. Belzig, and C. Bruder, Phys. Rev. A {\bf 68}, 043618 (2003).

\bibitem{haldane} F. D. M. Haldane, Phys. Rev. Lett. 47, 1840 (1981).

\bibitem{heiblum} E. Buks \etal, Nature {\bf 391}, 871 (1998).

\bibitem{stern} A. Stern, Y. Aharonov, and Y. Imry, Phys. Rev. A {\bf 41}, 3436(1990)

\bibitem{pad} A. Polkovnikov, E. Altman, and E. Demler, PNAS 103,
6125 (2006).
\bibitem{squeezed}
The interferometer can also be prepared in a
number squeezed initial state \cite{menotti,prentiss2}. This situation will not be
discussed in detail in the present paper.

\bibitem{polkovnikov} A. Polkovnikov, Phys. Rev. A, {\bf 68}, 033609 (2003); {\bf ibid.} {\bf 68}, 053604 (2003).

\bibitem{footnote} The interaction is activated using
$U(t)= \frac{1}{2}U(1+\tanh(\lambda t))$, where $\lambda$
controls the degree of adiabaticity.

\bibitem{cazalilla} M. A. Cazalilla, J. Phys. B 37, S1-S47 (2004).

\bibitem{footnote_xi} This length scale may be obtained by solving Gross-Pitaevskii
boundary problem in which the wave function increases from zero
at $x=0$ to its bulk value at $x \rightarrow \infty$. See e.g.  C. J. Pethick and H. Smith, "Bose-Einstein
condensation in dilute gases", p. 162, Cambridge University Press (2002).

\bibitem{olshanii} M. Olshanii, Phys. Rev. Lett. {\bf 81}, 938 (1998).

\bibitem{unpublished} R. Bistritzer and E. Altman, unpublished

\bibitem{gadp} V. Gritsev, E. Altman, and E. Demler, and A. Polkovnikov, cond-mat/0602475
\bibitem{jorg-private} J. Schmiedmayer, private communication
\bibitem{weiss} T. Kinoshita, T. Wenger, and D. S. Weiss, Nature {\bf 440}, 900 (2006).

\end{thebibliography}
\end{document}